\documentclass{aa}

\usepackage{graphicx}
\usepackage{txfonts}
\usepackage[colorlinks=true, linkcolor=blue, citecolor=blue, urlcolor=blue, breaklinks=true]{hyperref}
\begin{document} 

\title{Theoretical analysis of surface brightness--colour relations for late-type stars using MARCS model atmospheres}

\authorrunning{Salsi et al.}

\author{A. Salsi\inst{1}, N. Nardetto\inst{1}, B. Plez\inst{2},  D. Mourard\inst{1}}

   \institute{Université Côte d’Azur, OCA, CNRS, Laboratoire Lagrange, France\\
              \email{anthony.salsi@oca.eu}
              \and
    LUPM, Univ Montpellier, CNRS, Montpellier, France}

   \date{Received... ; accepted...}

% \abstract{}{}{}{}{} 
% 5 {} token are mandatory
  \abstract
   {Surface brightness--colour relations (SBCRs) are largely used for general studies in stellar astrophysics and  for determining extragalactic distances. 
   Based on a careful selection of stars and a homogeneous methodology, it has been recently shown  that the SBCR for late-type stars depends on the spectral type and luminosity class.}
   {Based on simulated spectra of late-type stars using MARCS model atmospheres, our aim is to analyse the effect of stellar fundamental parameters on the surface brightness. We also compare theoretical and recent empirical SBCRs.}
   {We used MARCS model atmospheres  to compute spectra and obtain the surface brightness of stars. We first explored the parameter space of MARCS (i.e. effective temperature, $\log g$,  $\mathrm{[Fe/H]}$, microturbulence, and mass) in order to quantify their impact on the surface brightness. Then we considered a relation between the effective temperature and $\log g$ for late dwarfs and giants, as well as a solar metallicity, in order to allow a consistent comparison of theoretical and empirical SBCRs.}
   {We find that the SBCR is not sensitive to the microturbulence and mass. The effect of metallicity on the SBCR is found to be larger for dwarfs than for giants. It is also larger when considering larger $V-K_s$ values. We also find  that a difference of 0.5 dex in metallicity between Galactic and LMC SBCRs does not affect the recent LMC distance determination, based on eclipsing binaries, by more than 0.4\%. By comparing theoretical with empirical SBCRs, we find a good agreement of less than 2$\sigma$ for F5--K7 dwarfs and giants stars, while a larger discrepancy is found for M dwarfs and giants (about 4--6$\sigma$). The surface gravity properties, as modelled in MARCS, explain the differences in the empirical SBCRs in terms of class. We finally find that theoretical and empirical SBCRs for Cepheids are consistent.}
   {Carefully considering metallicity and $\log g$ is mandatory when calibrating or using SBCRs.}
   

   \maketitle
   
%
%-------------------------------------------------------------------

\section{Introduction}\label{Intro}

Surface brightness--colour relations (SBCRs) are efficient tools for easily determining stellar angular diameters from photometric measurements. 
%Their ease of use makes these relations largely used in many international projects. 
In the course of the Araucaria project\footnote{\url{https://araucaria.camk.edu.pl/}}, \citet{2019Natur.567..200P}  estimated the distance to the Large Magellanic Cloud, based on 20 late-type eclipsing binaries, with a precision of 1\%. This achievement could be done using a precise SBCR calibrated on 41 nearby red clump giant stars \citep{2018AA...616A..68G}. In the same way, other works made use of giant late-type eclipsing binaries to constrain the Small Magellanic Cloud distance \citep{Graczyk2020}. To derive extragalactic distances from eclipsing binaries, the radii of the two components are estimated from the transit, by combining photometric and spectroscopic measurements. Then, the angular diameter of each component is estimated from the magnitude and colour of stars using a SBCR. Finally, the combination of radii (in kms) and angular diameters (in milliarcseconds) allows us to deduce the distance. The SBCRs should also play an important role in the context of the   PLAnetary Transits and Oscillations of stars (PLATO, \citealt{PLATO}) space mission, planned for launch in 2026, in order to characterise exoplanetary systems.

To date, authors have developed various SBCRs, covering all spectral types and luminosity classes. Several comparisons \citep{HDR_Nico} reveal precise but inconsistent SBCRs for late-type stars (i.e. $V-K_s > 1\,$ mag) at the 10\% level, while SBCRs for early-type stars (i.e. $V-K_s < 1\,$ mag) have recently been improved from around 7\% precision  \cite{2014AA...570A.104C} to 2--3\% \citep{2021AA...652A..26S}. Years ago, \cite{Fouque1997}  observed a significant difference in the SBCRs according to the luminosity class of stars. This dependance was then observed in several other studies \citep{Boyajian14, 2014AA...570A.104C, 2004AA...426..297K, HDR_Nico, vB99}. \cite{2020AA...640A...2S} (hereafter Paper~I)  calibrate SBCRs for late-type stars by implementing for the very first time criteria to properly select the samples, and use a homogeneous methodology in the calibration process. This allows us to clearly disentangle the SBCRs regarding their domain of spectral types (FGK and M stars) and luminosity classes. In \cite{2021AA...652A..26S} (hereafter Paper~II), we  convert our empirical SBCRs into uniform 2MASS-$K_s$ SBCRs.

In this paper we study theoretically the SBCRs for late-type stars (spectral types later than F5), in order to physically understand and try to reproduce the empirical 2MASS-$K_s$ SBCRs found in Paper~II.  We also restrict our analysis to the  $V - K_s$ colour.  We compute the surface brightness $F_V$ and the synthetic magnitudes $V$ and $K_s$ from models, and explore the impact of the model parameters on these quantities, and on the SBCRs.

We use MARCS stellar atmosphere models \citep{MARCS}. In Paper I, we  show that stellar activity, such as variability or fast rotation, but also multiplicity should be taken into account when calibrating and using the SBCRs to avoid any bias on the photometry. Stellar activity effects are not considered in our models, and the derived theoretical SBCR stands for inactive stars.

Section \ref{phot_section} details the surface brightness computation, as well as the photometric calibrations that are necessary  to compute the synthetic magnitudes from model spectra. Section~\ref{MARCS_section} is devoted to a description of the MARCS model atmospheres and the parameter space that we consider in the study. We present our results in Sect. \ref{section_influence}. Empirical and theoreticals SBCRs are compared in Sect. \ref{comparison_empirical}, and we discuss two aspects in Sect.~\ref{s_discussion}.

\section{Photometric calibration}\label{phot_section}

We used MARCS\footnote{Available at \url{https://marcs.astro.uu.se/}.} models \citep{MARCS} and, as in Paper I, focused our analysis on late-type stars,  stars with a spectral type later than F5. MARCS provides grids of one-dimensional, hydrostatic local thermodynamic equilibrium (LTE) model atmospheres in plane-parallel and spherical geometries \citep{MARCS}.  Various chemical composition classes are provided for atmospheric models, such as standard composition, $\alpha$-poor, $\alpha$-enhanced, or $\alpha$-negative. MARCS needs five atmospheric model parameters, namely the effective temperature $T_{\mathrm{eff}}$, the logarithmic surface gravity $\log g$, the overall metallicity $\mathrm{[Fe/H]}$, the microturbulence parameter $\mu$, and the mass $M_*$ for spherical geometry. The sampled flux (see \citealt{2008PhST..133a4003P}) is given in cgs units (ergs/cm$^2$/$\mathring{A}$), while $\lambda$ ranges from $0.13$ to $20\,\mu m$, sampled with a constant spectral resolution $\lambda/\Delta \lambda = 20.000$.

\subsection{Apparent surface brightness determination}

The surface brightness $F_V$ is defined as  \citep{SBCR_Wesselink}

\begin{equation}\label{def}
F_V = \log T_{\mathrm{eff}} + 0.1 BC_V,
\end{equation}where $BC_V$ is the visual bolometric correction. It is computed as $BC_V = M_{\mathrm{bol}} - M_V$, where $M_{\mathrm{bol}}$ is the bolometric magnitude of the star and $M_V$ its absolute magnitude. Once corrected from the  extinction $A_V$, the observed $V$ magnitude of a star at distance measured in parsec $d[\mathrm{pc}]$ can be expressed as

\begin{equation}
V = M_{\rm bol} - BC_V + 5\log d[\mathrm{pc}] -5, 
\end{equation}

which, using 

\begin{equation}
M_{\rm bol}  = -2.5 \log \left(\frac{L}{L_\odot}\right) + 4.74 \,\,\,\,\, {\rm and}\,\,\,\, L = 4\pi R^2\sigma T_{\rm eff}^4,
\end{equation}

where $M_{\rm bol}\odot = 4.74$ \citep{2015arXiv151006262M},
transforms into

\begin{multline}
V = -10 \log T_{\rm eff} -5 \log\left(\frac{R}{d[\mathrm{pc}]}\right) -0.26 -BC_V +5 \log R_\odot \\
+10 \log T_{\rm eff\odot}.
\end{multline}

Using $1 {\rm AU} = 149597870700\,{\rm m}$ \citep{2009CeMDA.103..365P}, the star angular diameter in milliarcseconds (mas) reads

\begin{equation}
\theta = \frac{1000}{149597870700} \frac{2 R}{d[\mathrm{pc}]},
\end{equation}

which allows us to write

\begin{equation}\label{sbcr}
F_V = -0.1 V - 0.5\log\theta + C,
\end{equation}

with

\begin{equation}
C = 0.5 \log \frac{2000}{149597870700} - 0.026 + 0.5 \log R_\odot + \log T_{\rm eff\odot}.
\end{equation}

To be consistent with  Paper I, we made use of solar constants from the IAU Resolution B2 \citep{2015arXiv151006262M}, and  we took the solar radius value from \cite{2018AA...616A..64M} (i.e. $T_{\mathrm{eff}\odot} = 5772\,$K, $R_\odot = 6.96134\times 10^8\,$m). This led to $C = 4.2196$, as was shown by \cite{2020AA...640A...2S}. This is slightly different from the original derivation of \cite{Barnes1976}. From Eq. \ref{sbcr}, the surface brightness appears as an apparent physical quantity, and therefore depends on the distance of the observer from the star. 

\subsection{Model surface brightness}

Computing  synthetic photometry with the filters defined in Sect. \ref{filters_sect} leads to magnitudes at the surface of the star which must be renormalised before comparison with the observed surface brightness. The model flux is $F_\lambda$, whereas the observed flux on earth is $f_\lambda = F_\lambda \left(\frac{R}{d}\right)^2$. The $V_{\rm mod}$ magnitude computed from the model flux is therefore 

\begin{equation}
V_{\rm mod} = V + 5 \log \left(\frac{R}{d}\right),
\end{equation}

or

\begin{equation}
V_{\rm mod} = V - 43.0773 - 5 \log \theta,
\end{equation}

which, inserted in Eq.\ref{sbcr}, gives the surface brightness for the model magnitudes

\begin{equation}
F_V = -0.1 (V_{\rm mod} + 43.0773) + 4.2196,
\end{equation}

or

\begin{equation}
    F_V = -0.1 V_{\mathrm{mod}} - 0.0881
.\end{equation}

\subsection{Filters and synthetic magnitudes}\label{filters_sect}

\begin{figure}
   \centering
   \includegraphics[width=\hsize]{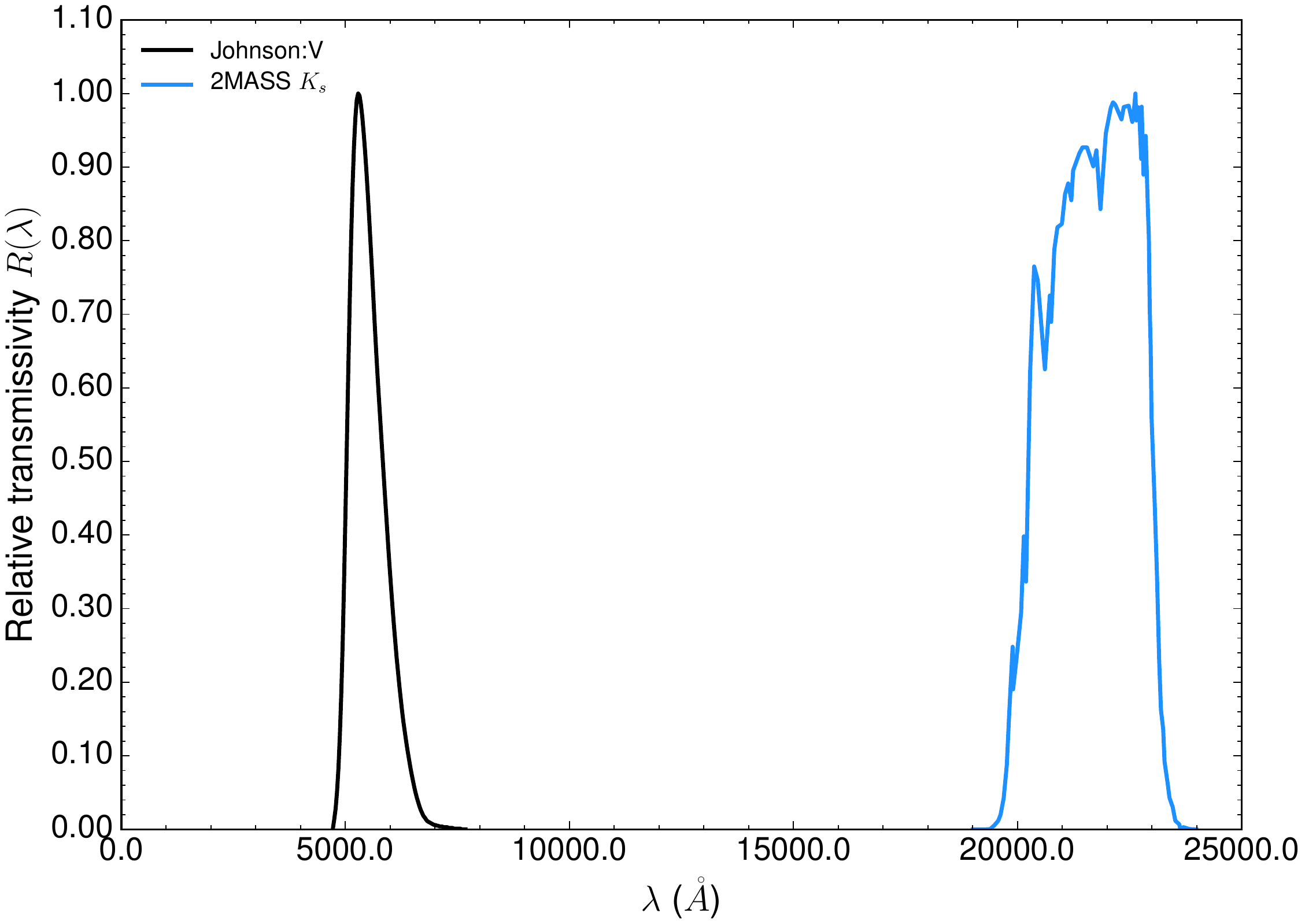}
   \caption{Relative Johnson:V (black  line) and 2MASS-$K_s$ (blue  line) transmissivity as a function of the wavelength $\lambda$.}
   \label{filters}
\end{figure}

To compute synthetic photometry, we recovered filters from the Spanish Virtual Observatory (SVO) Filter Profile service\footnote{\url{https://svo.cab.inta-csic.es/main/index.php}} \citep{SVO1, SVO2}. We considered the 2MASS filter described in \cite{2003AJ....126.1090C}. A large number of Johnson:V filters are available in the literature. We carefully selected a non-generic filter for consistency with the initial definition of Johnson:V magnitudes, with a large enough wavelength sampling leading to a smooth transmissivity curve. We followed the photometric calibration described in \cite{2018ApJS..236...47W}, and we thus used the recent Johnson:V filter recalibrated by \cite{2015PASP..127..102M}. The effective wavelength of this filter is $\lambda_{\mathrm{eff}} = 5452.41 \mathrm{\mathring{A}}$. The transmissivity of each filter is shown in Fig. \ref{filters}. By definition, the $V$ magnitude is computed using

\begin{equation}\label{calibration_mv}
    V = -2.5 \log \left[ \frac{\int f_V(\lambda) R_V(\lambda) \lambda d\lambda}{\int R_V(\lambda) \lambda d\lambda}\right] +zp,
\end{equation}

where $f_V(\lambda)$ is the stellar flux density, $R_V(\lambda)$ is the Johnson:V response function (i.e. the product of the detector quantum efficiency $\times$ filter throughput $\times$ unitless  fractional  transmission  of  the  total telescope optical train), and $zp$ is the zero-point correction. The integral is computed at each filter wavelength using a linear interpolation. %The flux that emerges from these spectra is the theoretical flux that would be measured at the surface of the star. 
The filter zero-point is adjusted, requesting that a standard star has the proper calculated
magnitudes. %The standard way to do the spectral calibration is to use the Vega's spectrum. %We discuss the impact of this choice later in the paper. 
For this purpose we used the STIS/CALSPEC Vega spectrum \citep{2014PASP..126..711B, 2020AAS...23537201B}.  The zero-point is found in such a way that the $V$ magnitude of Vega is 0.03$\,$mag. We find $zp = -21.09\,$mag, a value slightly different from previous studies \citep[i.e. -21.10$\,$mag;][]{1998AA...333..231B}. In the same way, we  calibrated the zero-point of the 2MASS filter to be -25.94$\,$mag, considering the convention $K_{s_{[\mathrm{Vega}]}} = 0\,$mag \citep{2MASS}. 

\section{Selecting MARCS model atmospheres}\label{MARCS_section}

At this date, MARCS contains more than 52,000 atmosphere models of late-type F, G, K, and M stars. The effective temperature of MARCS model atmospheres ranges from 2500 to 8000$\,$K, logarithmic surface gravities vary between -0.5 and 5.5, while overall logarithmic metallicities relative to the Sun $\mathrm{[Fe/H]}$ are  between -5.0 and +1.0. The stellar mass can be chosen between the standard mass of 1.0 solar mass and 15 solar masses in spherical geometry. Finally, the microturbulence parameter $\xi$ can be set at 0, 1, 2, or $5\,$km.s$^{-1}$ (see \citealt{MARCS} for more details).

In this work, we consider the standard composition \citep{2007SSRv..130..105G} to simulate spectra with $T_{\mathrm{eff}}$ between 2500$\,$K and 5000$\,$K, $\log g$ from -1 to 5, $\mathrm{\mathrm{[Fe/H]}}$ from $-2$ to $\mathrm{[Fe/H]} = 1$, and $\xi = 1, 2$, and $5\,$km.s$^{-1}$. For spherical models with $\log g <3$, we consider $M=1 M_{\odot}$ since we show in Sect.~\ref{sect_logg} that the impact of the mass on the SBCR is negligible.

\begin{figure*}
   \centering
   \includegraphics[width=0.5\hsize]{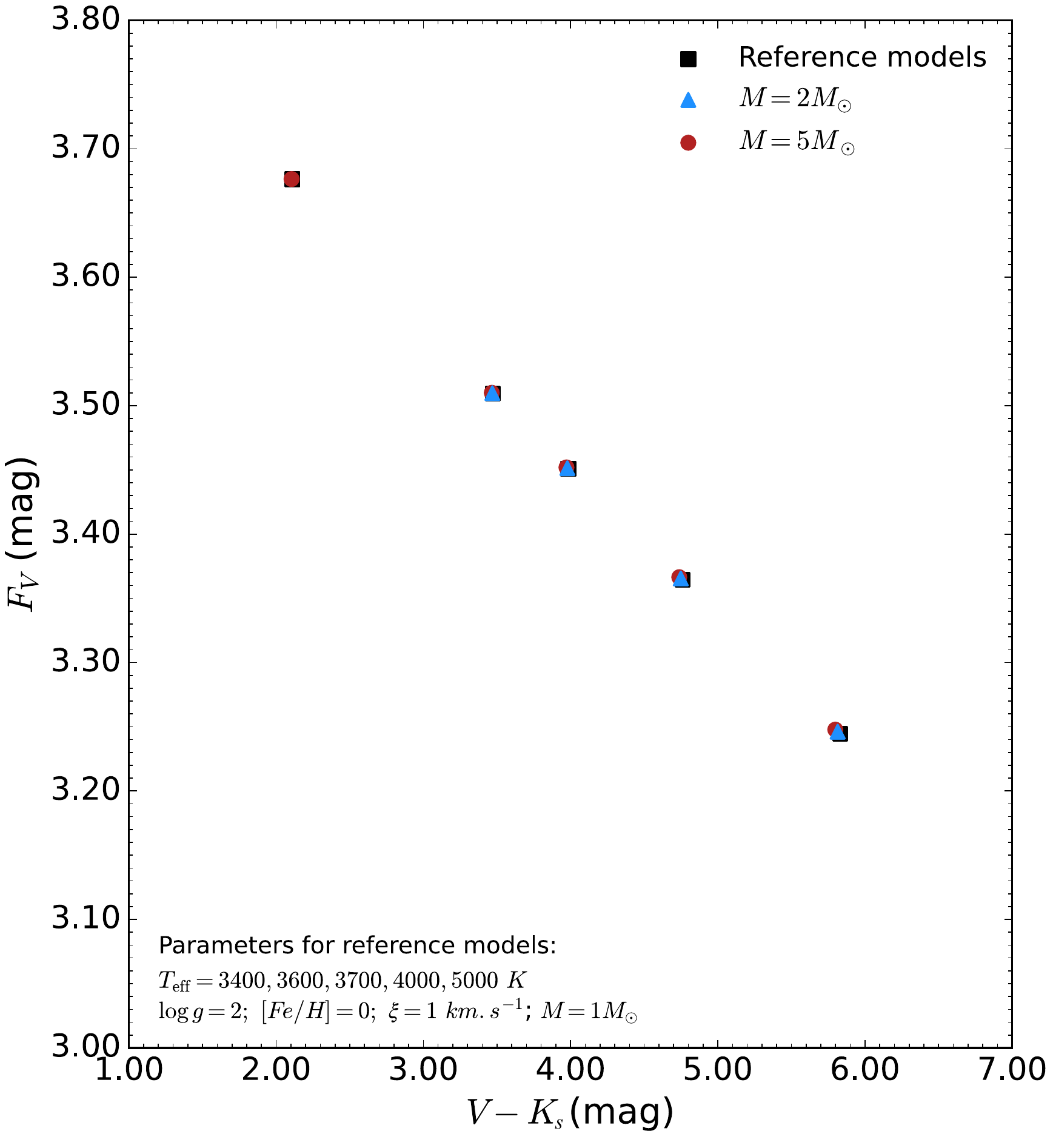}\hfill\includegraphics[width=0.5\hsize]{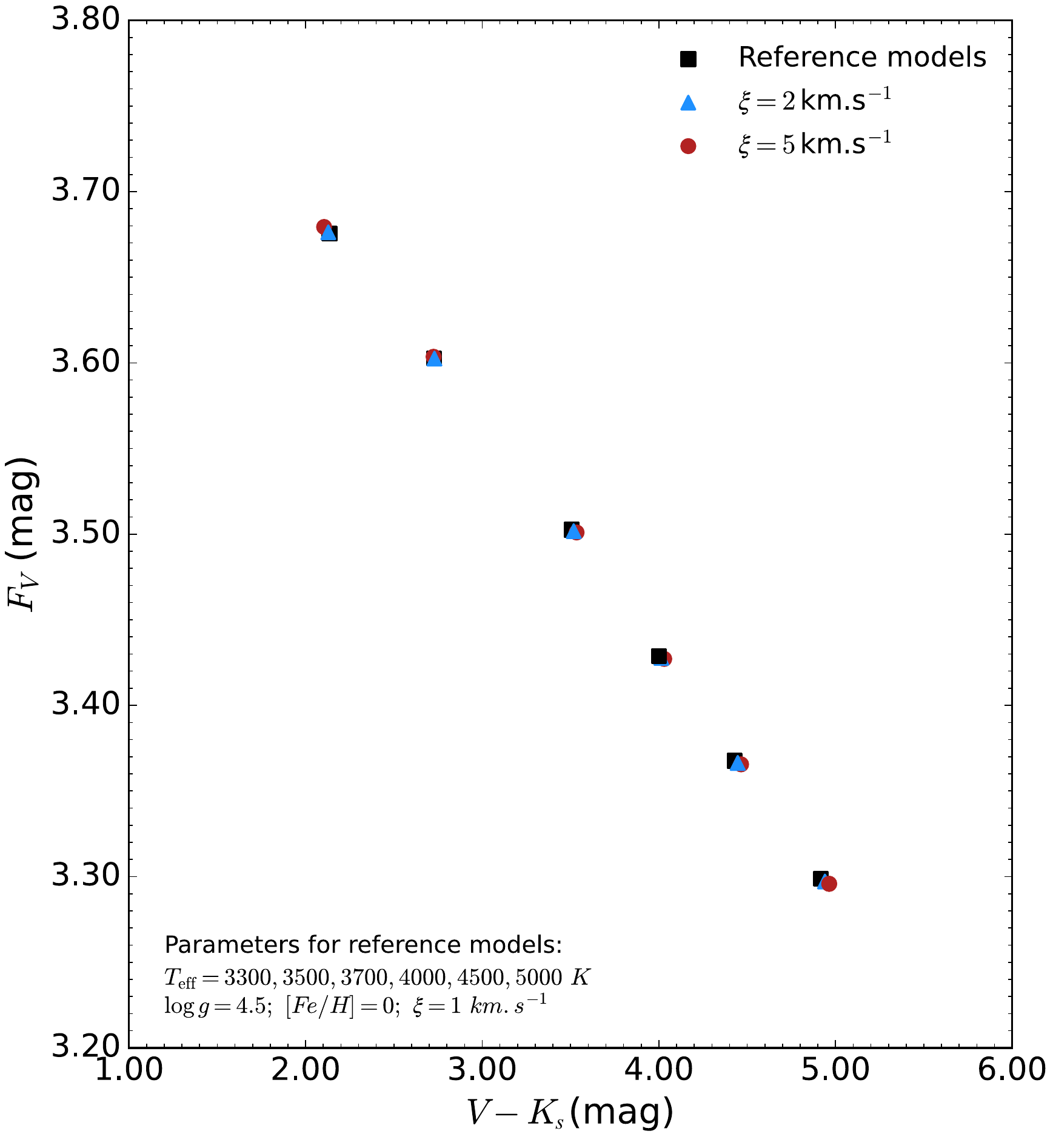}\\
   \includegraphics[width=0.5\hsize]{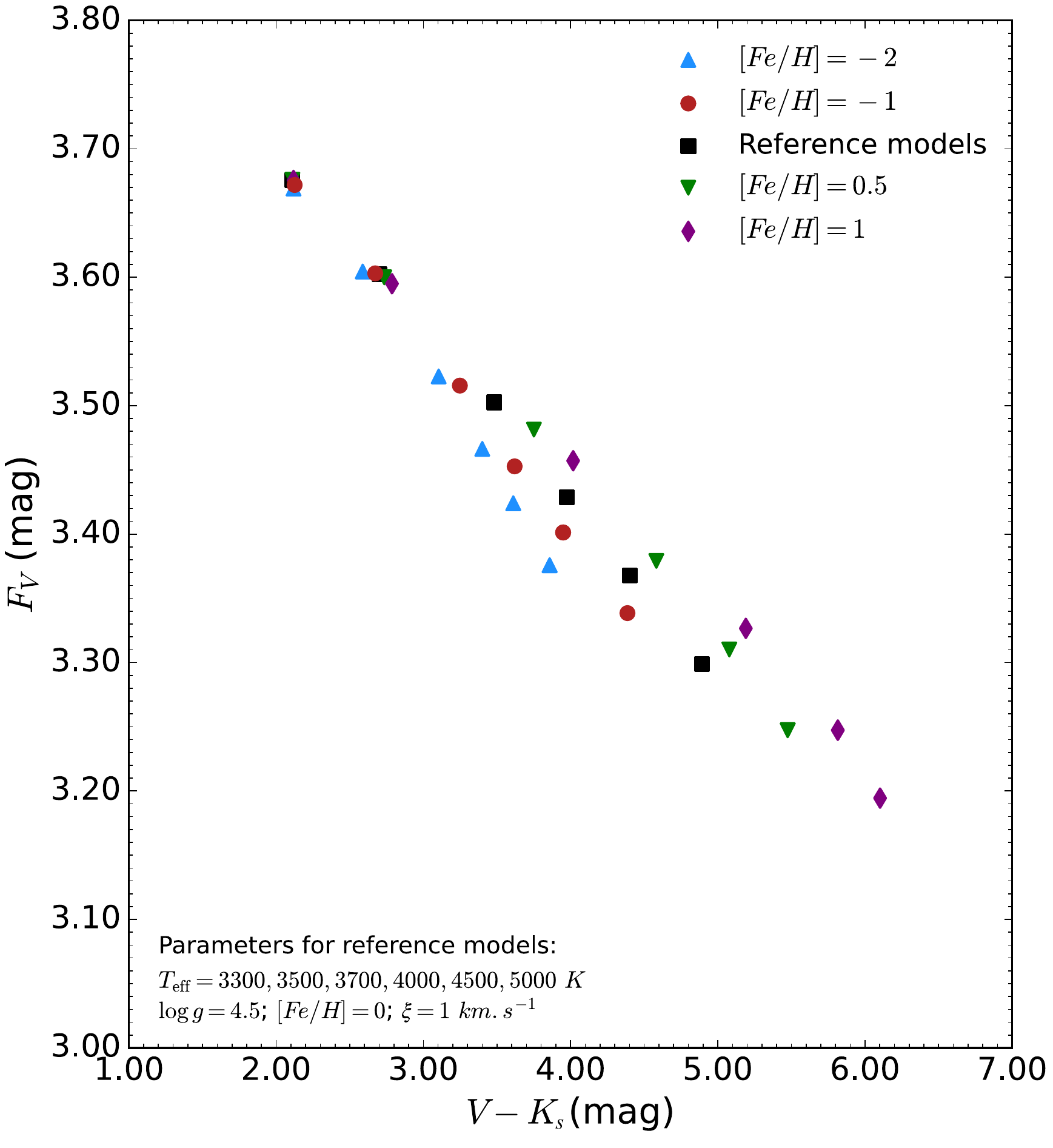}\hfill\includegraphics[width=0.5\hsize]{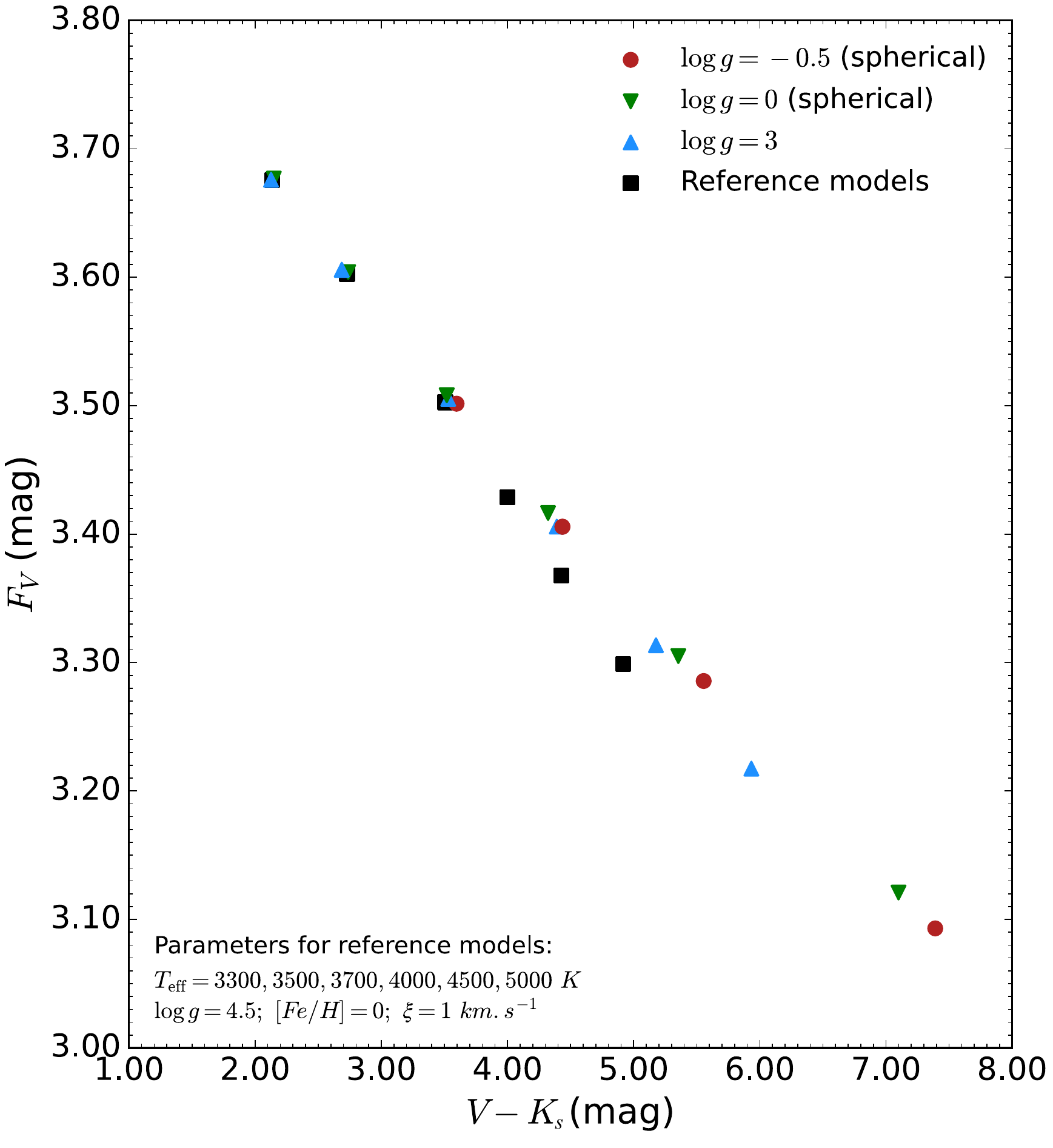}
   \caption{Influence of a change in fundamental parameters on the surface brightness of stars. Top left panel: Varying the stellar mass. Top right panel: Varying the microturbulence. Bottom left panel: Varying the metallicity. Bottom right panel: Varying the surface gravity.}
   \label{Standards_M}
\end{figure*}

\section{Impact of stellar model parameters on the SBCR}\label{section_influence}

\subsection{Reference atmosphere models}\label{reference_models}

We first define the reference models that   serve as elements of comparison. We consider $T_{\mathrm{eff}} =$ 3300, 3500, 3700, 4000, 4500, 5000$\,$K, and fix the other stellar parameters to a specific value depending on what we are studying.

\begin{figure*}
   \centering
   \includegraphics[width=0.498\hsize]{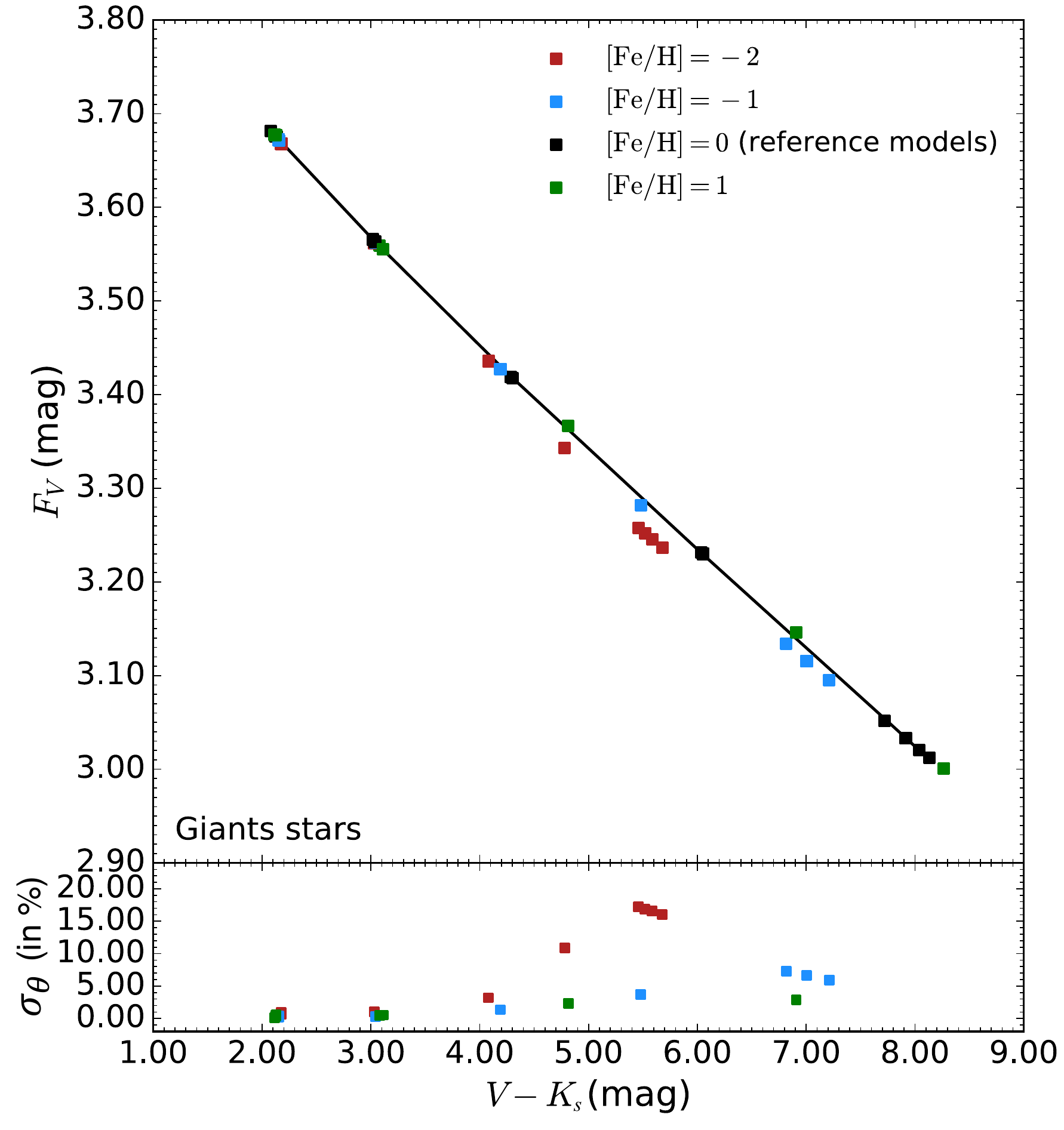}\hfill   \includegraphics[width=0.502\hsize]{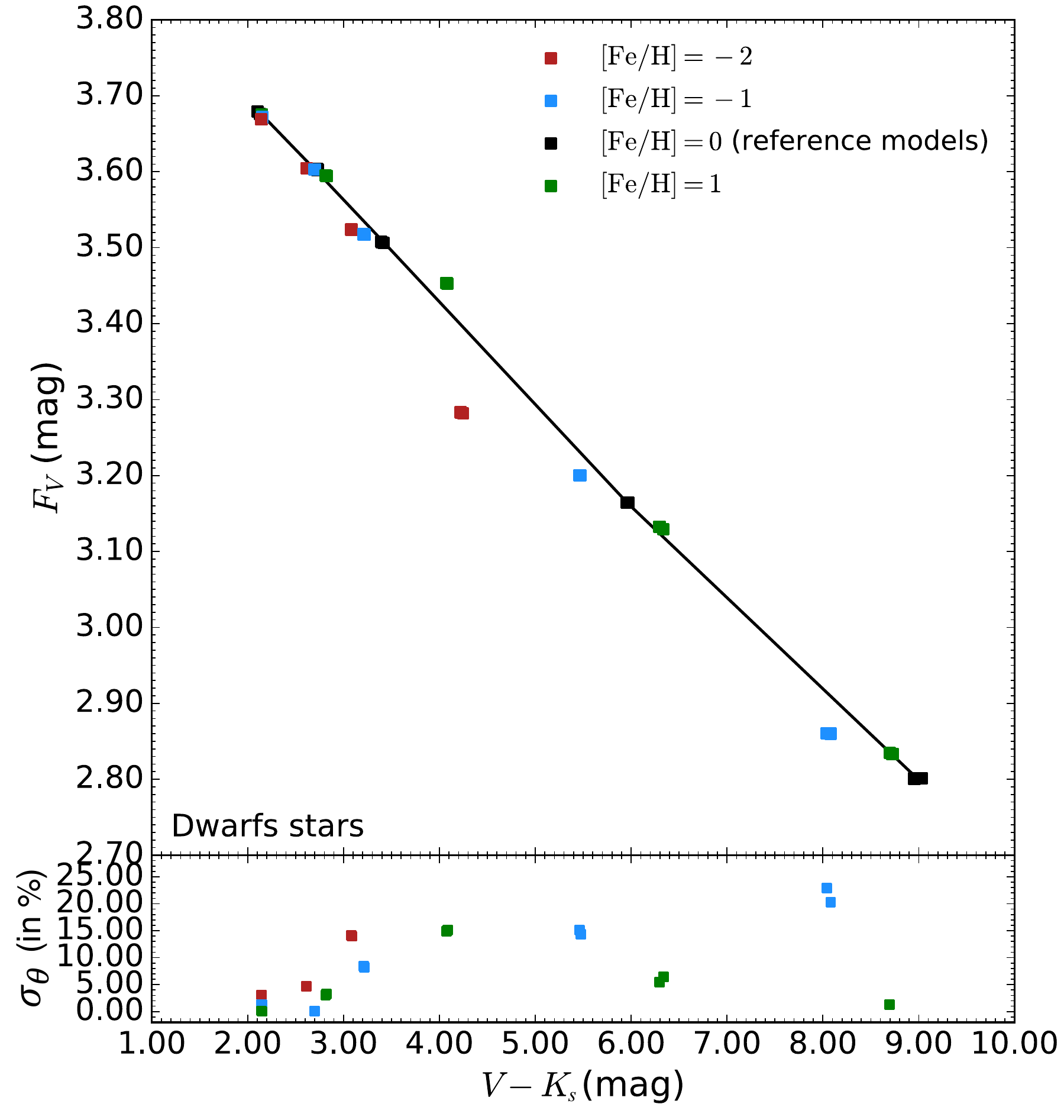}
   \caption{For giants (left) and dwarfs (right) the surface gravity $\log g$ varies with the effective temperature as shown in Table \ref{coeffs_logg} (black squares), while all the parameter space of MARCS is explored regarding the microturbulence and the mass. Other colours (red, blue, and green squares) correspond to different metallicities. The bottom panels show the absolute value of the difference in the expected angular diameter between a given SBCR and the reference value (solid black line).}
   \label{Standard_Z_SBCRs}
\end{figure*}

\subsection{Microturbulence}

We consider $\log g = 4.5$, $\mathrm{[Fe/H]} = 0$, and vary the microturbulence from $\xi = 1\,$km.s$^{-1}$ to $\xi = 5\,$km.s$^{-1}$. Results are shown in the  top right panel of Fig. \ref{Standards_M}. We see from these plots that both the surface brightness $F_V$ and the $V-K_s$ colour are only slightly sensitive to the microturbulence. In particular, both $F_V$ and $V-K_s$ values are shifted along the SBCR, which means that neither the slope nor the zero-point of the SBCR   depends on the microturbulence. We conclude that the SBCR does not depend on the  microturbulence of stars.

\begin{table}
\caption{Corresponding $\log g$ values to effective temperatures of giant (top) and dwarf (bottom) stars.}  
\centering                          % used for centering table
\begin{tabular}{ccc}        % centered columns (4 columns)

\hline\hline  
$(V-K_s)$       &       $T_{\mathrm{eff}}$      &       $\log g$ \\
$[\mathrm{mag}]$ & $[K]$ & \\
\hline          
\multicolumn{3}{c}{Giants} \\
\hline
2.1   & 5000 &  3.0 \\
3.0   & 4250 &  2.0 \\
4.3   & 3700 &  1.0 \\
6.0   & 3400 &  0.5 \\
8.0   & 3200 &  0.0 \\
\hline
\multicolumn{3}{c}{Dwarfs}\\
\hline
2.1   & 5000 &  4.5 \\
2.8   & 4500 &  4.5 \\
3.4   & 4000 &  5.0 \\
5.9   & 3000 &  5.0 \\
9.0   & 2500 &  5.0 \\
\hline
\end{tabular}
\label{coeffs_logg}   
\end{table}

\subsection{Stellar metallicity}\label{change_Fe}

To estimate the impact of the stellar metallicity $\mathrm{[Fe/H]}$, we simulate spectra with the following metallicities $\mathrm{[Fe/H]}$: -2, 1, 0, 0.5, and  1. The change in surface brightness of stars can be seen in the bottom left panel of Fig. \ref{Standards_M}. 

Increasing the metallicity leads to a decrease in the surface brightness and an increase in the colour of the star. This shift is negligibly small at $T_{\mathrm{eff}} \sim 5000\,$K and gradually increases at lower $T_{\mathrm{eff}}$. An offset of 0.025 magnitude in $F_\mathrm{V}$ (or 10\% on the angular diameter) is expected for a variation of one dex in $\mathrm{[Fe/H]}$ at $V-K_s \sim 4$.

\subsection{Stellar surface gravity}\label{sect_logg}

We studied the impact of stellar surface gravity on the surface brightness. We used models with $\log g$:  -0.5, 0, 3, and 4.5. In MARCS the models with $\log g < 3$ (resp. $\log g > 3$) are in spherical (resp. plane-parallel) geometry. To study the consistency of mixing different geometries, we  compared plane-parallel and spherical models with the same parameters. At $\log g = 3$ the difference is 0.05\% on the surface brightness of stars. Models in spherical geometry exist for different masses. In order to test this we considered the reference models described in Sect. \ref{reference_models} and set the mass to $M=2M_{\odot}$ and $M=5M_{\odot}$. We considered a value of $\log g = 2$. The result is shown in the top left panel of Fig. \ref{Standards_M}. We conclude that the impact of the mass is negligible in our study.

The bottom right panel of Fig. \ref{Standards_M} shows the influence of a change in surface gravity on the surface brightness deduced from atmosphere models, between $\log g = -0.5$ and $\log g = 4.5$. For $V-K_s$ larger than about 3
(i.e. $T_{\mathrm{eff}} = 4000\,$K)  both $F_V$ and the $V-K_s$ colour are strongly affected by $\log g$. For a star with $V-K_s = 4$, the difference in surface brightness ($F_\mathrm{V}$) for $\log g =$ -0.5, 0, and 3 is about 0.01 magnitude (or 5\% on the angular diameter), while it is significantly different for $\log g =$ 4.5 (about 0.025 magnitude or 10\% in angular diameter).

\begin{figure*}
   \centering
   \includegraphics[width=0.5\hsize]{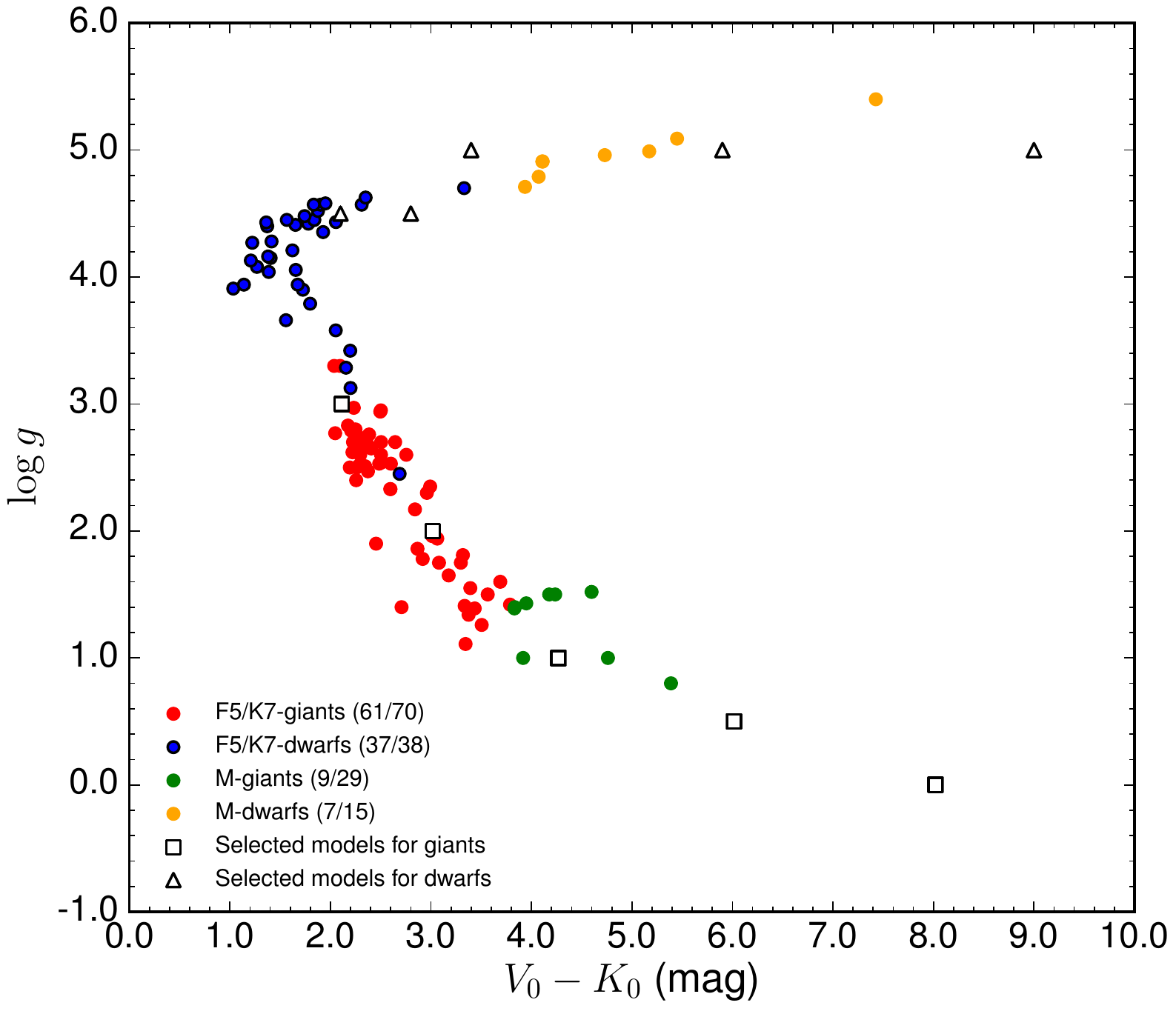}\hfill\includegraphics[width=0.5\hsize]{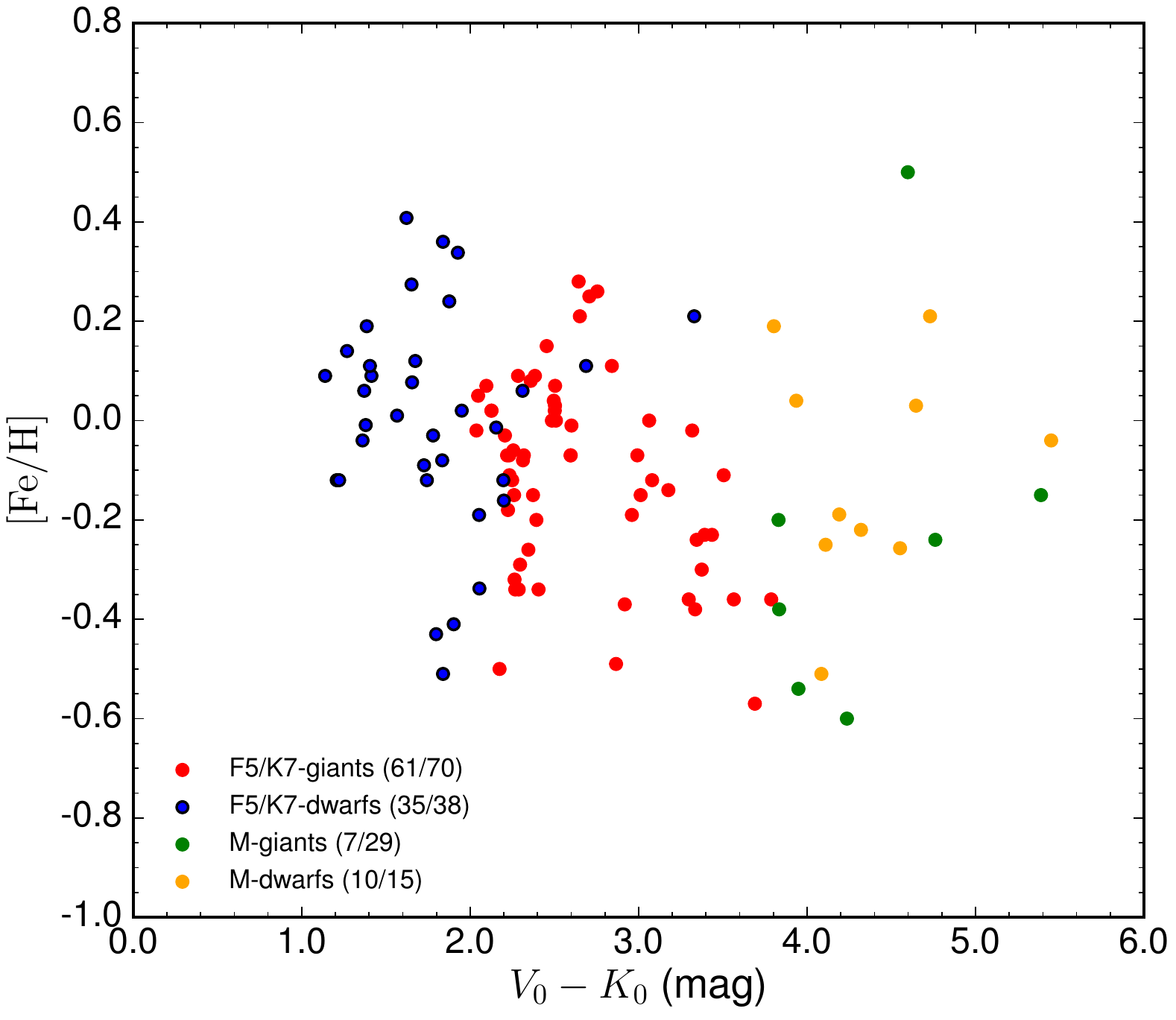}
   \caption{Fundamental parameters of stars in empirical samples of Paper I. Left panel: $\log g$ vs $V_0-K_0$ compared to coefficients from \cite{1998AA...333..231B} (see Table \ref{coeffs_logg}). Right panel: $\mathrm{[Fe/H]}$ vs $V_0-K_0$. The models used for the comparison of theoretical and empirical SBCRs have  solar metallicities  [Fe/H]=0.}
   \label{Fig_logg}
\end{figure*}

\begin{figure}
   \centering
   \includegraphics[width=\hsize]{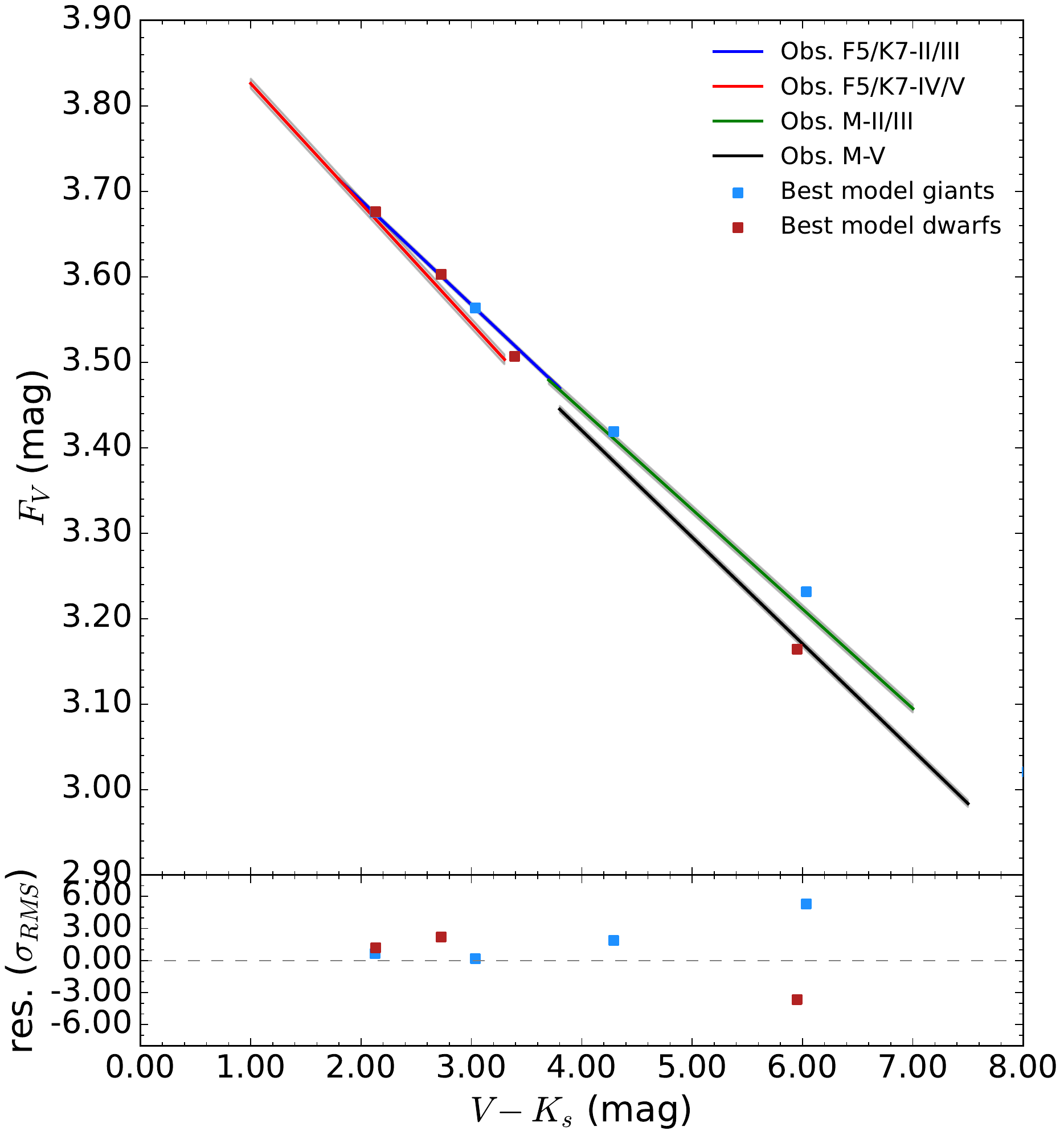}\hfill
   \caption{Comparison between 2MASS-$K_s$ SBCRs from Paper II and theoretical surface brightnesses calculated from stellar MARCS atmosphere models defined in Table \ref{coeffs_logg}, i.e. with a $\log g$ varying with the effective temperature. The metallicity is considered to be solar. The shaded grey area corresponds to the 1$\sigma$ confidence interval of the empirical SBCRs (corresponding in some cases to the thickness of the line). The lower panel shows the difference between the models and the corresponding SBCR (giants or dwarfs, respectively) in a fraction of the RMS of the empirical SBCR.}
   \label{comparison}
\end{figure}

Actually giants stars, but also dwarfs, have their surface gravity that varies with $T_{\mathrm{eff}}$. By considering standard evolution models, we can consider specific sets of models, as shown in Table \ref{coeffs_logg}  \citep{1998AA...333..231B}, that are also used for the comparison with observations in next section. In Fig. \ref{Standard_Z_SBCRs} we show the corresponding SBCRs (black line) in the case of giants (left) and dwarfs (right). We note that there are more than five references models (black squares) because we explore the entire parameter space  of MARCS models in term of microturbulence and mass. In this plot we also add the values of the surface brightness for different metallicities. For the sake of clarity, we do not consider the case of $\mathrm{[Fe/H]}=-0.5$. We find that the effect of metallicity on the SBCR is larger for dwarfs than for giants. It is also larger when considering larger $V-K_s$ values, and this is particularly true for metal-poor stars with [Fe/H]$=-2$ dex.

Thus, the theoretical analysis shows that the SBCR is insensitive to microturbulence. It is, however, sensitive to metallicity and stellar surface gravity. The effect of metallicity was already suggested by \cite{2004AA...426..297K} and \cite{2012ApJ...757..112B}, while the dependence of the SBCR with the class has already been observed through several works, such as \cite{Fouque1997, 2004AA...426..297K, G04} and \cite{Boyajian14}. However, no theoretical approach has been provided to date, except the one presented in \cite{2019PASP..131i4201M}. We come back to these results in the Conclusion.

\section{Comparison of theoretical and empirical SBCRs}\label{comparison_empirical}
 
We performed a very first comparison of observations from paper I with the MARCS models. In Paper I, we   implement four precise SBCRs (converted into uniform 2MASS-$K_s$ SBCRs in Paper II) for F5--K7 giants, F5--K7 subgiants or dwarfs, M giants, and M dwarfs stars. We  cross-matched the empirical samples from Paper I with the various references in the literature (using SIMBAD\footnote{\url{http://simbad.u-strasbg.fr/simbad/sim-fbasic}} queries) in order to recover the surface gravity and the metallicity. 

Over 152 stars, we found 114 values of $\log g$ and 113 values of $\mathrm{[Fe/H]}$ (see Table~\ref{Tab_logg}). We plot the $\log g$ and $\mathrm{[Fe/H]}$ values as a function of $V_0-K_0$ in Fig.~\ref{Fig_logg}.

We find that the stars used to calibrate the four SBCRs have solar metallicities on average, with a standard deviation of at most 0.5 dex in [Fe/H], which basically corresponds to the step in the MARCS grid. For $\log g$, we see that our reference models as indicated in Table \ref{coeffs_logg} (i.e. open squares for giants and open triangles for dwarfs), are consistent with observations.

Finally, for a consistent comparison, we consider the set of models of Table~\ref{coeffs_logg} with  solar metallicity. The microturbulence and the mass, as shown previously, have little impact on the surface brigthness and are set to $0$ km.s$^{-1}$ and 1 solar mass, respectively. Two models are rejected in Table \ref{Tab_logg} because they exceed the $V-K_s$ validity domain of the empirical relations. We end up with eight models for the comparison.\\ The results are shown in Fig.~\ref{comparison} and can be summarised as follows. The empirical 2MASS-$K_s$ SBCRs for F5--K7 stars (dwarfs or giants) are systematically 1--2$\sigma$ brighter than the theoretical ones. As the RMS of the empirical relations are of 0.004 and 0.002 magnitude in $F_\mathrm{V}$ respectively for dwarfs and giants, this difference corresponds to 0.002--0.008 magnitude in $F_\mathrm{V}$ or a 1--4\% at most in terms of angular diameter.
 For M stars (dwarfs or giants), the difference between the observations and the MARCS models is larger, between 5 and 6$\sigma$. Interestingly, the empirical SBCR for giants is brighter than the theoretical one, while it is the contrary for the dwarfs. The RMS of the empirical relations are of 0.004 and 0.005 magnitude in $F_\mathrm{V}$ respectively for dwarfs and giants. This difference corresponds to around 10\% in terms of angular diameter.
%\end{itemize}

In this analysis, it is not excluded that the zero-points of the theoretical SBCRs are affected by the filters and/or the reference star used for the calculation of the synthetic magnitudes. We used Vega as a reference star, while it is known to be a pole-on fast rotator \citep{2006ApJ...645..664A}. 

If we use instead the STIS/CALSPEC Sirius spectrum \citep{2020AAS...23537201B} to calculate the photometric zero-point, we find  -21.12$\,$mag instead of -21.09$\,$mag. This offset of 0.03 magnitude on the reference star leads to an offset of 0.003 magnitude on $F_\mathrm{V}$, which corresponds to a 1--1.5 RMS of the empirical SBCRs. Using $\eta$ UMa as reference, considering its STIS/CALSPEC spectrum \citep{2020AAS...23537201B}, leads to a zero-point of -21.07$\,$mag, corresponding to a difference of $\sim$0.002 mag on $F_\mathrm{V}$. Such offsets only partially explain the difference observed when comparing empirical and theoretical SBCRs. Interestingly, there is one single existing filter for the 2MASS-$K_s$ magnitude. This is a strong advantage compared to the various Johnson $K$ filters we can find in the literature. By choosing the 2MASS-$K_s$ photometry in this study, we excluded any bias that could be induced by the choice of the filter.

Another possible bias can be the filter from which the synthetic $m_V$ magnitude is computed. The stellar flux is  integrated over a wavelength range and a transmissivity that are both specific to the filter. We made a test by using the generic Johnson:V filter of \cite{1998AA...333..231B}. With this filter, the zero-point is found to be -21.12$\,$mag. We observe a constant difference of 0.006$\,$mag on $F_\mathrm{V}$ with respect to the \cite{2015PASP..127..102M} filter. A change in the filter and/or in the reference star can lead to a better agreement between zero-points of empirical and theoretical relations, but cannot fully explain the differences obtained, in particular for M stars.

\section{Discussion}\label{s_discussion}

\subsection{Effect of the metallicity on the LMC distance}

\begin{figure}
   \centering
   \includegraphics[width=\hsize]{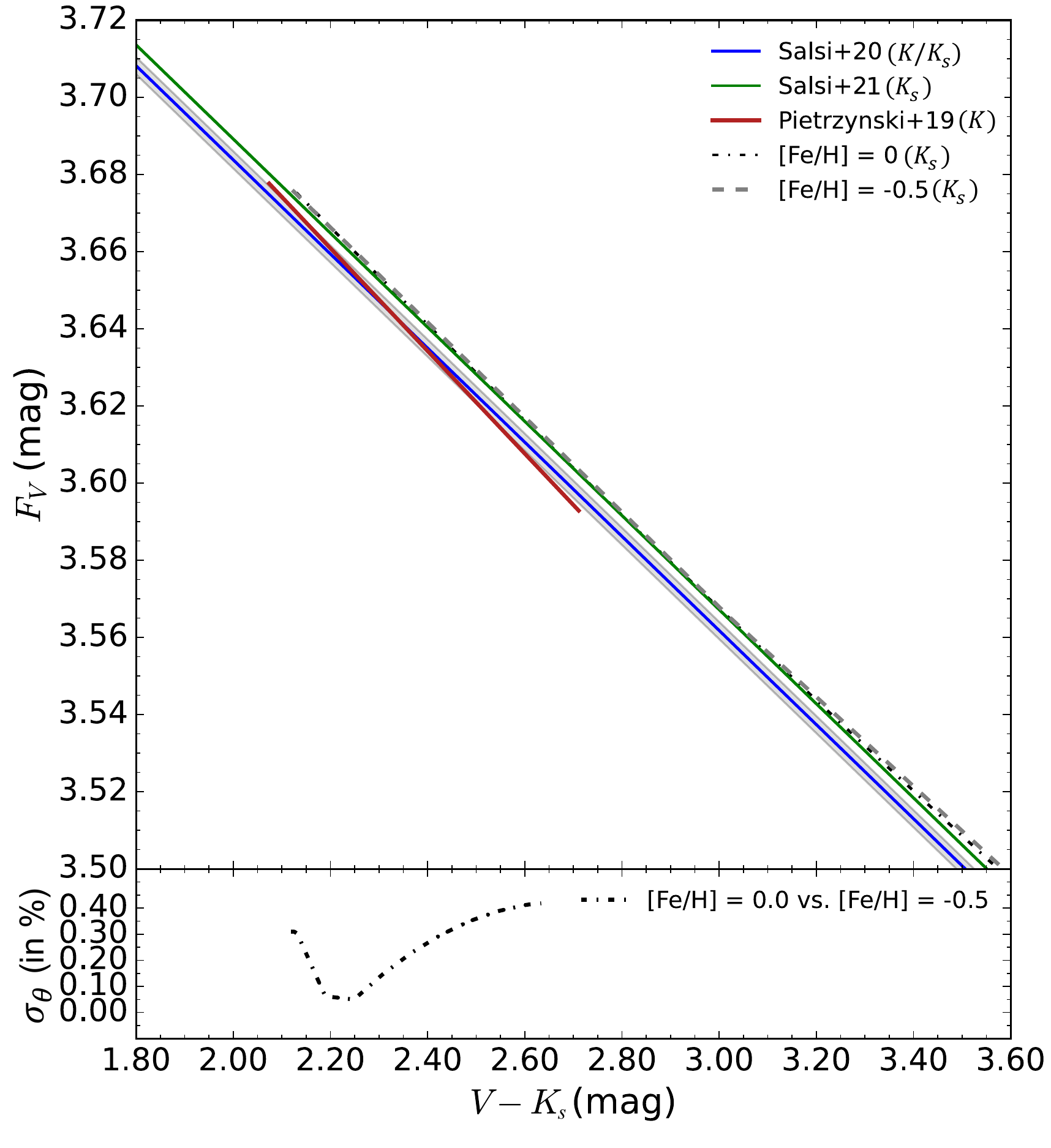}\hfill
   \caption{Comparison of theoretical SBCRs based on atmosphere models with metallicity of $\mathrm{[Fe/H]} = 0.0$ and $\mathrm{[Fe/H]} = -0.5$, respectively. The empirical SBCRs from Paper I, Paper II, and \cite{2019Natur.567..200P} are overplotted for comparison. The difference in the derived angular diameter $\theta_\mathrm{LD}$ (in \%) using the two theoretical relations of different metallicity is shown in the bottom panel.}
   \label{comparison_distance}
\end{figure}

Recently, \cite{2019Natur.567..200P} have established the distance to LMC with a precision of 1\% using a SBCR based on 41 Galactic red clump giant stars \citep{2018AA...616A..68G}. 
%We showed above that the metallicity has a strong impact on the calibration of SBCRs. It is interesting to evaluate at which level the distance of the LMC can be impacted by the metallicity of stars using SBCRs. 
The metallicity of the stars ranges from -0.66 to 0.34 dex (see \cite{2018AA...616A..68G}, with an average of $-0.01 \pm 0.07$ dex, while the metallicity of the LMC is of about -0.4 dex \citep{2016MNRAS.455.1855C, 2021MNRAS.507.4752C}.
To test the impact of metallicity on the SBCR and the distance of LMC, we compare theoretical SBCRs with metallicities of $\mathrm{[Fe/H]} = 0.0$ and $\mathrm{[Fe/H]} = -0.5$ in Fig. \ref{comparison_distance}. The difference in the derived angular diameter ($\theta_\mathrm{LD}$) using both SBCRs is less than 0.4\% over the colour domain of validity of the \cite{2019Natur.567..200P} relation, with an average of about 0.25\% (see lower panel of Fig. \ref{comparison_distance}). This basically means that a decrease in $\mathrm{[Fe/H]}$ of 0.5 dex on the SBCR leads to an increase in the LMC distance of at most 0.25\% (or 0.25$\sigma$ when considering the precision of the distance of LMC established by \citealt{2019Natur.567..200P}). In the figure we also show for comparison the empirical SBCRs of \cite{2019Natur.567..200P}, and of Paper I and   Paper II. These SBCRs are consistent, but they are slightly shifted to brighter surface brightenesses compared to the theoretical SBCRs. 

\subsection{Theoretical SBCRs for Cepheids}

\begin{figure}
   \centering
   \includegraphics[width=\hsize]{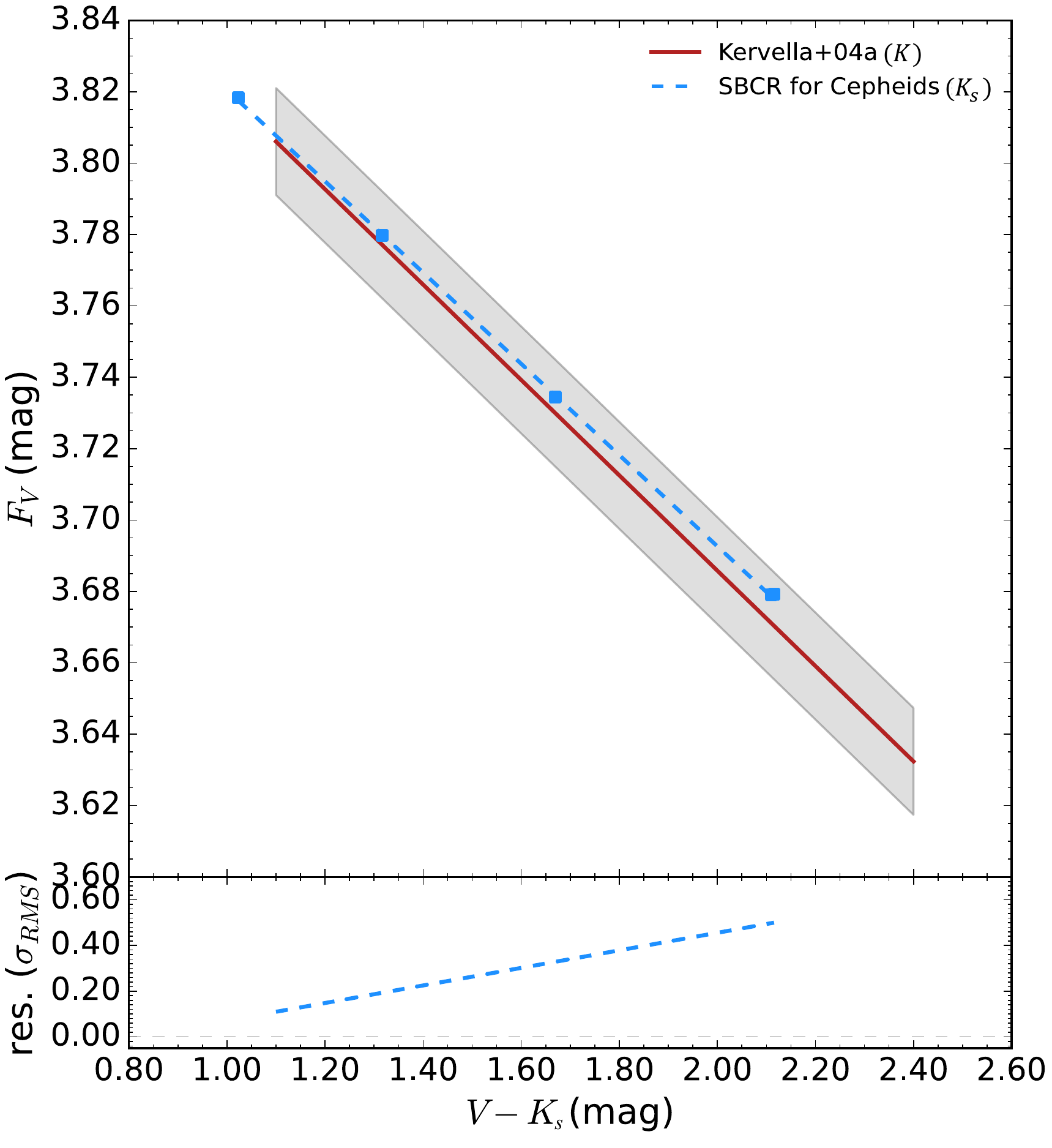}\hfill
   \caption{Comparison between the theoretical SBCR for Cepheids and the empirical relation from \cite{2004AA...428..587K}. The lower panel shows the difference in fraction of RMS.}
   \label{comparison_Ceph}
\end{figure}

The period-luminosity relation of Cepheids \citep{2020AA...643A.115B} is used to calibrate the Hubble--Lemaitre constant $H_0$ \citep{2016ApJ...826...56R, 2021arXiv211204510R}. However, the different versions of the  Baade--Wesselink (BW) method of distance determination, based on a SBCR \citep{2011AA...534A..94S, 2011AA...534A..95S}, interferometric observations \citep{2004AA...416..941K}, or even both \citep{2021arXiv211109125T}, are currently not used to calibrate $H_0$, mainly because of the projection factor \citep{2004AA...428..131N, 2017AA...597A..73N} and circumstellar environment issues \citep{2020AA...633A..47H, 2021AA...651A..92H, 2021AA...651A.113G}. In this context, the calibration of the SBCR of Cepheids is crucial in order to better understand the physics of Cepheids. Our  aim here is to compare the SBCR usually used for the application of the BW (i.e. \citealt{2004AA...428..587K}), and the theoretical one. For this, we consider several atmosphere models within the Cepheid instability strip \citep{2021arXiv211109125T}, as indicated in Table \ref{coeffs_logg_Cepheids}. We consider a solar metallicity of  $\mathrm{[Fe/H]} = 0$. In Fig. \ref{comparison_Ceph} we show a comparison between the theoretical SBCR for Cepheids and the empirical SBCR of \cite{2004AA...428..587K}. The difference is lower than 0.3$\sigma$ over the whole validity domain. This result shows an excellent agreement between theoretical and empirical SBCRs for Cepheids.

\begin{table}
\caption{Corresponding $\log g$ values to effective temperatures of Cepheid stars.}
\centering                          % used for centering table
\begin{tabular}{ccc}        % centered columns (4 columns)
\hline\hline  
$(V-K_s)$       &       $T_{\mathrm{eff}}$      &       $\log g$ \\
$[\mathrm{mag}]$ & $[K]$ & \\
\hline          
\multicolumn{3}{c}{Cepheids} \\
\hline
1.0  & 6500 &  2.5 \\
1.3   & 6000 &  2.0 \\
1.6   & 5500 &  1.5 \\
2.1   & 5000 &  1.0 \\
2.1   & 5000 &  0.5 \\
\hline
\end{tabular}
\label{coeffs_logg_Cepheids}   
\end{table}

\section{Conclusion and perspectives}

In Paper I we  showed that empirical SBCRs are dependant on the luminosity class of stars. By using the MARCS model atmospheres in this paper, we  have theoretically analysed the influence of fundamental stellar parameters on the surface brightness of stars. We confirm the result of Paper I, and we show that SBCRs vary with the surface gravity of stars, and therefore depend on the luminosity class. Though the effect of mass and microturbulence is weak, we have shown that the metallicity also impacts the SBCR. The metallicity should therefore be taken into consideration when calibrating and using a SBCR. In this respect, we  show that a difference in metallicity of 0.5 dex in the calibration of SBCRs does not impact the LMC distance by more than 0.4\% (with an average difference of 0.25\%).

For the first time, we have compared empirical and theoretical SBCRs. We find a very good agreement of about 1--2$\sigma$ level for F5--K7 stars, while it is of 5--6$\sigma$ for M stars. Such discrepancies, on the theoretical side, could be partly due to the choice of the reference star and/or the filter used in the models. On the observation side, it is not excluded that some stars, in particular M giants stars, are affected by dust environment, which might alter the colour estimate. 

Finally, by comparing the theoretical SBCR for Cepheids with the empirical one used in the BW method of distance determination \citep{2004AA...428..587K} we find an excellent agreement, better than 0.3$\sigma$.

\bibliographystyle{aa}
\bibliography{salsi}

\begin{acknowledgements}
      BP acknowledges support from the CNES (Centre National d’Etudes Spatiales) in the framework of the PLATO mission. This work is based upon observations obtained with the Georgia State University Center for High Angular Resolution Astronomy Array at Mount Wilson Observatory. The CHARA Array is funded by the National Science Foundation through NSF grants AST-0606958 and AST-0908253 and by Georgia State University through the College of Arts and Sciences, as well as the W. M. Keck Foundation. This work made use of the JMMC Measured stellar Diameters Catalog \citep{Duvert}. This research made use of the SIMBAD and VIZIER\footnote{Available at \url{http://cdsweb.u- strasbg.fr/}} databases at CDS, Strasbourg (France) and the electronic bibliography maintained by the NASA/ADS system. This work has made use of data from the European Space Agency (ESA) mission Gaia (\url{https://www.cosmos.esa.int/gaia}). This research also made use of Astropy, a community-developed core Python package for Astronomy \citep{astropy2018}. This research has made use of the SVO Filter Profile Service\footnote{\url{http://svo2.cab.inta-csic.es/theory/fps/}} supported from the Spanish MINECO through grant AYA2017-84089.
\end{acknowledgements}

\clearpage
\onecolumn

\begin{appendix}

\section{Additional table}

\begin{longtable}{lccccl}
\caption{{Characteristics of empirical samples from Paper I.}}\label{Tab_logg}\\
\hline\hline
Star HD &       Box     &       $V_0-K_0$       &       $\log g$        &       $\mathrm{[Fe/H]}$       &       Source  \\
 & & [mag] & & & \\
\hline
\endfirsthead
\caption{Continued.}\\
\endhead
\endfoot
\hline
HD 10142        &       1       &       2.37    &       2.47    &       -0.15   &       \cite{2015MNRAS.448.2749A}      \\
HD 102328       &       1       &       2.76    &       2.60    &       0.26    &       \cite{2019AA...625A.141L}       \\
HD 113226       &       1       &       2.05    &       2.77    &       0.05    &       \cite{2018ApJS..238...29P}      \\
HD 11977        &       1       &       2.19    &       2.50    &       -       &       \cite{2016AA...585A...5B}       \\
HD 120477       &       1       &       3.69    &       1.60    &       -0.57   &       \cite{2007AA...475.1003H}       \\
HD 127665       &       1       &       2.96    &       2.30    &       -0.19   &       \cite{2007AA...475.1003H}       \\
HD 133124       &       1       &       3.51    &       1.26    &       -0.11   &       \cite{2019AA...625A.141L}       \\
HD 13468        &       1       &       2.25    &       2.80    &       -0.12   &       \cite{2007AA...475.1003H}       \\
HD 135722       &       1       &       2.27    &       2.63    &       -0.34   &       \cite{2019AA...625A.141L}       \\
HD 136726       &       1       &       3.06    &       1.94    &       0.00    &       \cite{2016AA...588A..98M}       \\
HD 153210       &       1       &       2.50    &       2.70    &       0.07    &       \cite{2007AA...475.1003H}       \\
HD 157681       &       1       &       3.34    &       1.11    &       -0.24   &       \cite{2019AA...625A.141L}       \\
HD 163770       &       1       &       2.71    &       1.40    &       0.25    &       \cite{1990AJ.....99.1961F}      \\
HD 164058       &       1       &       3.39    &       1.55    &       -0.23   &       \cite{1981ApJ...248..228L}      \\
HD 16815        &       1       &       2.41    &       2.65    &       -0.34   &       \cite{2015MNRAS.448.2749A}      \\
HD 170693       &       1       &       2.87    &       1.86    &       -0.49   &       \cite{2019AA...625A.141L}       \\
HD 176678       &       1       &       2.50    &       2.95    &       0.02    &       \cite{2007AA...475.1003H}       \\
HD 17709        &       1       &       3.79    &       1.42    &       -0.36   &       \cite{1990ApJS...74.1075M}      \\
HD 17824        &       1       &       2.10    &       3.30    &       0.07    &       \cite{2007AA...475.1003H}       \\
HD 184293       &       1       &       2.92    &       1.78    &       -0.37   &       \cite{2019AA...625A.141L}       \\
HD 185958       &       1       &       2.21    &       2.79    &       -0.03   &       \cite{1990ApJS...74.1075M}      \\
HD 18784        &       1       &       2.38    &       2.76    &       0.09    &       \cite{2001ApJ...551L..85Z}      \\
HD 192781       &       1       &       3.44    &       1.39    &       -0.23   &       \cite{2019AA...625A.141L}       \\
HD 19787        &       1       &       2.28    &       2.69    &       0.09    &       \cite{2019AA...627A.138A}       \\
HD 200205       &       1       &       3.34    &       1.41    &       -0.38   &       \cite{2019AA...625A.141L}       \\
HD 204381       &       1       &       2.04    &       3.30    &       -0.02   &       \cite{2007AA...475.1003H}       \\
HD 211388       &       1       &       3.08    &       1.75    &       -0.12   &       \cite{1990ApJS...74.1075M}      \\
HD 214868       &       1       &       3.01    &       1.96    &       -0.15   &       \cite{2019AA...625A.141L}       \\
HD 215665       &       1       &       2.26    &       2.40    &       -0.06   &       \cite{2012AA...543A.160T}       \\
HD 216131       &       1       &       2.13    &       2.99    &       0.02    &       \cite{2018AA...615A..31D}       \\
HD 216131       &       1       &       2.13    &       2.99    &       0.02    &       \cite{2018AA...615A..31D}       \\
HD 219449       &       1       &       2.49    &       2.53    &       0.00    &       \cite{2019AA...625A.141L}       \\
HD 220572       &       1       &       2.36    &       2.73    &       0.08    &       \cite{2007MNRAS.382..553L}      \\
HD 23526        &       1       &       2.26    &       2.50    &       -0.15   &       \cite{2014ApJ...785...94L}      \\
HD 23940        &       1       &       2.29    &       2.52    &       -0.34   &       \cite{2015MNRAS.448.2749A}      \\
HD 30504        &       1       &       3.30    &       1.75    &       -0.36   &       \cite{1990ApJS...74.1075M}      \\
HD 30814        &       1       &       2.23    &       2.97    &       -0.07   &       \cite{1990ApJS...74.1075M}      \\
HD 3546 &       1       &       2.17    &       2.83    &       -0.50   &       \cite{2015AA...580A..24D}       \\
HD 360  &       1       &       2.32    &       2.73    &       -0.07   &       \cite{2007MNRAS.382..553L}      \\
HD 36848        &       1       &       2.64    &       2.70    &       0.28    &       \cite{2006AA...458..609D}       \\
HD 36874        &       1       &       2.51    &       2.54    &       0.00    &       \cite{2011AA...536A..71J}       \\
HD 3750 &       1       &       2.50    &       2.60    &       0.03    &       \cite{2007MNRAS.382..553L}      \\
HD 39523        &       1       &       2.45    &       1.90    &       0.15    &       \cite{1984PhDT........85P}      \\
HD 39640        &       1       &       2.23    &       2.70    &       -0.11   &       \cite{2015MNRAS.448.2749A}      \\
HD 4211 &       1       &       2.60    &       2.53    &       -0.01   &       \cite{2007MNRAS.382..553L}      \\
HD 46116        &       1       &       2.26    &       2.63    &       -0.32   &       \cite{2015MNRAS.448.2749A}      \\
HD 5722 &       1       &       2.22    &       2.70    &       -0.18   &       \cite{2007AA...475.1003H}       \\
HD 60060        &       1       &       2.31    &       2.72    &       -0.08   &       \cite{2015MNRAS.448.2749A}      \\
HD 60341        &       1       &       2.50    &       2.94    &       0.04    &       \cite{2015AA...580A..24D}       \\
HD 69267        &       1       &       3.37    &       1.34    &       -0.30   &       \cite{2018AA...620A..58S}       \\
HD 76294        &       1       &       2.22    &       2.62    &       -0.07   &       \cite{2019AA...625A.141L}       \\
HD 83618        &       1       &       2.99    &       2.35    &       -0.07   &       \cite{2007AA...475.1003H}       \\
HD 85503        &       1       &       2.65    &       -       &       0.21    &       \cite{2019MNRAS.490.1821C}      \\
HD 8651 &       1       &       2.39    &       2.66    &       -0.20   &       \cite{2015MNRAS.448.2749A}      \\
HD 87837        &       1       &       3.32    &       1.81    &       -0.02   &       \cite{1990ApJS...74.1075M}      \\
HD 9362 &       1       &       2.30    &       2.60    &       -0.29   &       \cite{2015MNRAS.448.2749A}      \\
HD 9408 &       1       &       2.35    &       2.51    &       -0.26   &       \cite{2015AA...580A..24D}       \\
HD 96833        &       1       &       2.60    &       2.33    &       -0.07   &       \cite{2019AA...625A.141L}       \\
HD 96833        &       1       &       2.60    &       2.33    &       -0.07   &       \cite{2019AA...625A.141L}       \\
HD 98262        &       1       &       3.18    &       1.65    &       -0.14   &       \cite{2019AA...625A.141L}       \\
HD 9927 &       1       &       2.84    &       2.17    &       0.11    &       \cite{2016AA...588A..98M}       \\
HD 99998        &       1       &       3.57    &       1.50    &       -0.36   &       \cite{2019AA...627A.138A}       \\
HD 102870       &       2       &       1.27    &       4.08    &       0.14    &       \cite{2017AJ....153...21L}      \\
HD 10476        &       2       &       1.95    &       4.58    &       0.02    &       \cite{2017AJ....153...21L}      \\
HD 10697        &       2       &       1.67    &       3.94    &       0.12    &       \cite{2017AJ....153...21L}      \\
HD 10700        &       2       &       1.84    &       4.45    &       -0.51   &       \cite{2019AA...629A..80H}       \\
HD 114710       &       2       &       1.37    &       4.40    &       0.06    &       \cite{2017AJ....153...21L}      \\
HD 117176       &       2       &       1.73    &       3.90    &       -0.09   &       \cite{2017AJ....153...21L}      \\
HD 140283       &       2       &       1.56    &       3.66    &       -2.43   &       \cite{2019AA...627A.138A}       \\
HD 140538       &       2       &       1.57    &       4.45    &       0.01    &       \cite{2017AJ....153...21L}      \\
HD 142860       &       2       &       1.21    &       4.13    &       -0.12   &       \cite{2017AJ....153...21L}      \\
HD 158633       &       2       &       1.90    &       4.57    &       -0.41   &       \cite{2017AJ....153...21L}      \\
HD 16160        &       2       &       2.35    &       4.63    &       -       &       \cite{2019AA...629A..80H}       \\
HD 168723       &       2       &       2.20    &       3.13    &       -0.16   &       \cite{2019AA...629A..80H}       \\
HD 173667       &       2       &       1.14    &       3.94    &       0.09    &       \cite{2017AJ....153...21L}      \\
HD 173701       &       2       &       1.84    &       4.45    &       0.36    &       \cite{2018MNRAS.481.3244G}      \\
HD 175726       &       2       &       1.36    &       4.43    &       -0.04   &       \cite{2017AJ....153...21L}      \\
HD 182572       &       2       &       1.62    &       4.21    &       0.41    &       \cite{2019AA...629A..80H}       \\
HD 185144       &       2       &       1.83    &       4.57    &       -0.08   &       \cite{2017AJ....153...21L}      \\
HD 187637       &       2       &       1.22    &       4.27    &       -0.12   &       \cite{2018ApJ...861..149F}      \\
HD 188512       &       2       &       2.05    &       3.58    &       -0.19   &       \cite{2017AJ....153...21L}      \\
HD 188887       &       2       &       2.69    &       2.45    &       0.11    &       \cite{2007MNRAS.382..553L}      \\
HD 190360       &       2       &       1.65    &       4.41    &       0.27    &       \cite{2019AA...629A..80H}       \\
HD 19373        &       2       &       1.41    &       4.15    &       0.11    &       \cite{2017AJ....153...21L}      \\
HD 195564       &       2       &       1.66    &       4.06    &       0.08    &       \cite{2019AA...629A..80H}       \\
HD 198149       &       2       &       2.20    &       3.42    &       -0.12   &       \cite{2017AJ....153...21L}      \\
HD 19994        &       2       &       1.39    &       4.04    &       0.19    &       \cite{2017AJ....153...21L}      \\
HD 21019        &       2       &       1.80    &       3.79    &       -0.43   &       \cite{2017AJ....153...21L}      \\
HD 219134       &       2       &       2.31    &       4.57    &       0.06    &       \cite{2018ApJS..238...29P}      \\
HD 22484        &       2       &       1.38    &       4.16    &       -0.01   &       \cite{2019AA...629A..80H}       \\
HD 30652        &       2       &       1.03    &       3.91    &       -       &       \cite{2019AA...629A..80H}       \\
HD 34411        &       2       &       1.41    &       4.28    &       0.09    &       \cite{2017AJ....153...21L}      \\
HD 3651 &       2       &       1.88    &       4.52    &       0.24    &       \cite{2017AJ....153...21L}      \\
HD 38858        &       2       &       1.74    &       4.48    &       -0.12   &       \cite{2017AJ....153...21L}      \\
HD 4628 &       2       &       2.06    &       4.43    &       -0.34   &       \cite{2019AA...629A..80H}       \\
HD 69830        &       2       &       1.78    &       4.42    &       -0.03   &       \cite{2019AA...629A..80H}       \\
HD 75732        &       2       &       1.93    &       4.35    &       0.34    &       \cite{2019AA...629A..80H}       \\
HD 88230        &       2       &       3.33    &       4.70    &       0.21    &       \cite{2017AJ....153...21L}      \\
HD 90043        &       2       &       2.15    &       3.29    &       -0.01   &       \cite{2019AA...629A..80H}       \\
HD 1013 &       3       &       4.18    &       1.50    &       -       &       \cite{2008AJ....135..209M}      \\
HD 102212       &       3       &       3.95    &       1.43    &       -0.54   &       \cite{2019AA...627A.138A}       \\
HD 120933       &       3       &       4.60    &       1.52    &       0.50    &       \cite{1990ApJS...74.1075M}      \\
HD 121130       &       3       &       4.76    &       1.00    &       -0.24   &       \cite{1986ApJ...311..843S}      \\
HD 183439       &       3       &       3.83    &       1.40    &       -0.38   &       \cite{1986ApJ...311..843S}      \\
HD 18884        &       3       &       4.24    &       1.50    &       -0.60   &       \cite{2012AA...547A.108L}       \\
HD 19058        &       3       &       5.39    &       0.80    &       -0.15   &       \cite{1986ApJ...311..843S}      \\
HD 218329       &       3       &       3.83    &       1.39    &       -0.20   &       \cite{2016AA...587A...2B}       \\
HD 25025        &       3       &       3.92    &       1.00    &       -       &       \cite{2008AJ....135..209M}      \\
GJ406   &       4       &       7.43    &       5.40    &       -       &       \cite{2018AA...620A.180R}       \\
GJ447   &       4       &       5.45    &       5.09    &       -0.04   &       \cite{2019AA...625A..68S}       \\
GJ581   &       4       &       4.73    &       4.96    &       0.21    &       \cite{2018ApJS..238...29P}      \\
GJ674   &       4       &       4.55    &       -       &       -0.26   &       \cite{2019AA...629A..80H}       \\
GJ687   &       4       &       4.65    &       -       &       0.03    &       \cite{2014MNRAS.443.2561G}      \\
GJ876   &       4       &       5.17    &       4.99    &       -       &       \cite{2019AA...625A..68S}       \\
HD 119850       &       4       &       4.07    &       4.79    &       -       &       \cite{2019AA...627A.138A}       \\
HD 1326 &       4       &       4.11    &       4.91    &       -0.25   &       \cite{2019AA...625A..68S}       \\
HD 199305       &       4       &       3.94    &       4.71    &       0.04    &       \cite{2019AA...625A..68S}       \\
HD 204961       &       4       &       4.19    &       -       &       -0.19   &       \cite{2019AA...629A..80H}       \\
HD 225213       &       4       &       4.09    &       -       &       -0.51   &       \cite{2019AA...629A..80H}       \\
HD 36395        &       4       &       3.80    &       -       &       0.19    &       \cite{2019AA...629A..80H}       \\
HIP51397        &       4       &       4.32    &       -       &       -0.22   &       \cite{2019AA...629A..80H}       \\
\hline
\end{longtable}
\tablefoot{From left to right:  Star HD,  boxes relative to Paper I (1: F5--K7 II--III, 2: F5--K7 IV--V, 3: M II--III, 4: M V),  $V-K_s$ colour corrected from the interstellar extinction,  logarithmic surface gravity $\log g$,  metallicity $\mathrm{[Fe/H]}$,  reference of the $\log g$, and $\mathrm{[Fe/H]}$ measurements.}

\end{appendix}

\end{document}